# GÖRÜNTÜ İŞLEME VE YAPAY SİNİR AĞLARI İLE İLETİM HATLARINDA ARIZA YERİ BELİRLEME

[1]**Serkan BUDAK**, [2]**Bahadır AKBAL**

[1]*Konya Teknik Üniversitesi, Mühendislik ve Doğa Bilimleri Fakültesi, Elektrik-Elektronik Mühendisliği Bölümü, KONYA*
[2]*Konya Teknik Üniversitesi, Mühendislik ve Doğa Bilimleri Fakültesi, Elektrik-Elektronik Mühendisliği Bölümü, KONYA*
[1]sbudak@ktun.edu.tr, [2]bakbal@ktun.edu.tr,

**ÖZ:** Elektrik enerjisinin kesintisiz ve kaliteli bir şekilde iletilmesi için, üretim yapıldığı noktadan tüketim olan noktaya kadar kontrol edilmesi gerekmektedir. Dolayısıyla üretimden tüketime kadar her aşamada iletim ve dağıtım hatlarında koruma yapılması şarttır. Elektrik tesislerinde koruma rölelerinin temel görevi, sistemde meydana gelen kısa devrelerde arızalı olan bölgenin mümkün olan en kısa sürede devre dışı etmektir. Sistemin en önemli parçası olan enerji iletim hatları ve bu hatları koruyan mesafe koruma rölelerine bu konuda çok önemli görevler düşmektedir. Hızlı ve verimli çalışmalar yapmak için doğru bir hata yeri tespit tekniği gereklidir. İletim hatlarında transformatör nötr nokta topraklaması bir güç sisteminin tek faz – toprak kısa devre arızası sırasında oluşan sıfır bileşen akımı mesafe koruma rölesinin çalışmasını etkilemektedir. Topraklama sistemi ve koruma sistemleri arasındaki ilişki göz önüne alındığında, uygun bir topraklama seçimi yapılmalıdır. İletim hatlarında farklı topraklama sistemlerinde kısa devre arızalarının yerinin doğru bir şekilde belirlenebilmesi için yapay sinir ağı (YSA) kullanılmıştır. YSA'nın performansını test etmek için destek vektör makineleri (DVM) ile karşılaştırılmıştır. İletim hattı modeli PSCAD ™ / EMTDC ™ benzetim programında oluşturulup YSA için gerekli veriler elde edilmiştir. Farklı topraklama sistemlerinde oluşturulan kısa devre arızalarındaki mesafe koruma rölesinin R-X empedans diyagramının empedans değişiminin görüntüsü kayıt altına alınarak veri setleri oluşturulmuştur. Görüntülerde ilgili odak noktaları özellik çıkarım ve görüntü işleme teknikleri kullanılarak farklı YSA modellerine giriş olarak verilmiş ve en iyi arıza yeri tahmini veren YSA modeli seçilmiştir.

***Anahtar Kelimeler:*** *Mesafe Koruma Rölesi, Arıza Yeri Tahmini, Görüntü İşleme Teknolojisi, YSA, Kısa Devre Arızaları, Güç Transformatör Topraklaması*

# Determination of Fault Location in Transmission Lines with İmage Processing and Artificial Neural Networks

**ABSTRACT:** In order to transmit electrical energy in a continuous and quality manner, it is necessary to control it from the point of production to the point of consumption. Therefore, protection of transmission and distribution lines is essential at every stage from production to consumption. The main function of the protection relays in electrical installations should be deactivated as soon as possible in the event of short circuits in the system. The most important part of the system is energy transmission lines and distance protection relays that protect these lines. An accurate error location technique is required to make fast and efficient work. Transformer neutral point grounding in transmission lines affects the operation of the zero component current during the single phase to ground short circuit failure of a power system. Considering the relationship between the grounding system and protection systems, an appropriate grounding choice should be made. Artificial neural network (ANN) has been used in order to accurately locate short circuit faults in different grounding systems in transmission lines. Compared with support vector machines (SVM) for testing inside ANN The transmission line model is made in the PSCAD ™ / EMTDC ™ simulation program. Data sets were created by recording the image of the impedance change of the R-X impedance diagram of the distance protection relay in short circuit faults created in different grounding systems. The related focal points in the images are given as an introduction to different ANN models using feature extraction and image processing techniques and the ANN model with the highest fault location estimation accuracy was chosen.

## GİRİŞ (INTRODUCTION)

Modern güç sistemlerinde elektrik enerjisinin üretilmesinden, yerleşim birimlerine iletilmesine kadar olan süreçte sürekliliğin sağlanması çok önemlidir. Elektrik tesislerinde koruma rölelerinin temel fonksiyonu, sistemde meydana gelen kısa devrelerde ya da tesise ve tesis elemanlarına zarar verebilecek herhangi bir anormal durumda, elektrik şebekesinin geri kalan kısmının etkin bir şekilde çalışmasını etkilemeyecek şekilde, arızalı olan bölgenin mümkün olan en kısa sürede devre dışı edilmesini sağlamaktır.

Kısa devrelerin elektrik tesislerindeki etkileri çok farklı olabilmektedir. Darbe şeklinde ani olarak ortaya çıkması ile oluşan büyük kısa devre akımları, tesis elemanları üzerinde hatlara, kablolara, baralara, transformatörlere ve diğer donanımlara mekanik olarak zarar verebilir. Devreden uzun süre geçen sürekli kısa devre akımları ise tesis elemanlarının ısınmasına ve malzemenin termik bakımdan zorlanmasına sebep olabilir. Bundan dolayı hem tesis hem de işletme personeli bundan zarar görebilir. Kısa devre olayının sebep olduğu arızalar sonucunda, işletme kısmen veya tamamen durur ve kademe kademe enerji üretimi, iletimi, dağıtımı ve tüketimi normal olarak devam edemez. Ayrıca arızanın sebep olduğu hasarlar büyük onarım masraflarına yol açar.

Elektrik sistemindeki arızaların çok önemli bir kısmı dengesiz olup, çok az bir kısmı dengelidir. Arızaların pratikte oluşma sıklığı, tesisatın yapısına, çevreye ve bölgeye göre değişse de sıralama genelde benzerdir.

Bir araştırmaya göre bu sıklık;
- Üç fazlı simetrik kısa devre için %5,
- İki faz-toprak kısa devresi için %10,
- Faz-faz kısa devresi için %15,
- Faz-toprak kısa devresi için %70,

olarak belirlenmiştir (Grainger ve diğ., 2003).

Sistemin en önemli parçası olan enerji iletim hatları ve bu hatları koruyan mesafe koruma rölelerine bu konuda çok önemli görevler düşmektedir. Mesafe koruma röleleri, iletim ve dağıtım hatlarında ana ve yedek koruma olarak geniş bir kullanıma sahiptirler. İletim hatlarını korumak için en çok tercih edilen röledir (Glover ve diğ., 2012). Temel olarak mesafe koruma röleleri gerilim ve akım değerlerini karşılaştırarak hattın empedansına göre arızayı ve arızanın yerini belirlemektedir (Ziegler, 2011).

Modern güç sistemlerinin uygun teknikle, ekonomik ve düzenleyici yapıda güvenilir şekilde çalışması gerekmektedir. Bu nedenle koruma özelliği, iletim sistemlerinde daha iyi kontrol edilebilecek ve daha verimli, daha güvenilir hizmet kalitesi sağlayacak şekilde gözden geçirilmelidir.

İletim hattında bir arıza meydana geldiğinde, mesafe röleleri arızalı hattı, arızanın türünü ve arıza yerini algılar (Ram ve diğ., 2013). Mesafe koruma rölesi, farklı arıza durumlarından, dalga formlarındaki DC ofsetinin (simetrik olmayan akım), sıfır bileşen akımı ve ağ yapılandırma değişikliklerinden etkilenebilir. (Jihong ve diğ., 1993), (Liao ve Elangovan, 1998), (Ye ve diğ., 1998) karmaşık hesaplamalar içeren empedansa dayalı yöntemlerde çoğu zaman kısa devre arıza yeri tahmininde hata oluşturur.

Yürüyen dalga tabanlı yöntemlerde sinyal işleme metodu da bağlı olarak doğru arıza ve arıza yeri tahmini yapılabilir, fakat akım transformatörleri kısa devre anında doyuma gitmeden ve doğru sonuçlar vermesi gerekmektedir. Bu makale YSA kullanarak yukarıdaki zorlukların üstesinden gelebilecek yeni bir yaklaşım sunmaktadır.

Literatürde, yapılan çalışmalarda mesafe koruma rölesi arıza ve arıza yeri tespitinde genel olarak empedansa dayalı yöntemler, yürüyen dalga tabanlı yöntemler ve sezgisel yöntemler kullanılmaktadır. (Yağan, 2015) Dağıtım hatlarında hat başından alınan üç faza ait gerilim ve akım bilgileri dijital arıza kayıt edicide işlenerek belirli frekans gruplarına ayrılarak YSA'ya verilmiştir. (Liang ve Jeyasurya, 2004),(Maheshwari ve diğ., 2019), (Osman ve Malik, 2004), (Zubić ve diğ., 2017), (Jung ve diğ., 2007) İletim hatlarının mesafe koruması için dalgacık dönüşümü uygulaması kullanılmıştır. Özellikle, dijital röle tarafından ölçülen gerilim ve akım sinyallerinin temel frekans bileşenlerini çıkarmak için kullanılır. (Dos Santos ve Senger, 2011), (Chawla ve diğ., 2006) sezgisel tabanlı algoritma kullanarak, röleden arıza noktasına hat empedansını tahmin eden bir fonksiyon tahmincisi olarak çalıştırmıştır ve geleneksel mesafe koruma rölelerine benzer şekilde, bu algoritma sırasıyla akım ve gerilim transformatörleri tarafından sağlanan akım ve gerilim örneklerine dayalı olarak hat empedansını tahmin eden ölçüm sonuçlarına dayanmaktadır. (Zhong ve diğ., 2013) Faz-toprak arızaları için üç katsayılı doğrusal bir diferansiyel denklem çözülerek ölçüm noktasından arıza noktasına arıza mesafesini hesaplayan yeni bir mesafe koruma algoritması önerilmiştir. (Ray ve Mishra, 2016) uzun bir iletim hattındaki destek vektörü makine tabanlı hata tipi ve mesafe tahmin şemasını önermektedir. Önerilen teknik, arıza sonrası tek çevrim akım dalga formunu kullanır ve örneklerin ön işleme dalgacık dönüşümü ile yapılır. Dalgacık dönüşümü ile çok sayıda özellik toplanması, fazlalık özelliklerin kullanılmaması için ileri özellik seçim yönteminin uygulanması, böylece tahmin doğruluğunun artırılması sağlamaktadır. (Swetapadma ve Yadav, 2018) paralel hatlarda her türlü arızanın arıza yeri tahmini için k-en yakın komşu (k-NN) tabanlı yöntem önerilmektedir. Ayrık Fourier Dönüşümü (DFT), sinyallerin ön-işlenmesi için kullanılır ve daha sonra bir ön-arıza çevriminin ve bir arıza-sonrası numune döngüsünün standart sapması k-NN algoritmasına giriş olarak kullanılır. (Meddeb ve diğ., 2019) sıfır bileşen akımlarının, mesafe rölesi korumasının çalışmasının etkisini önlemek için yeni bir yaklaşım önermektedir. Önerilen yöntem, tek fazlı bir arıza sırasında güç transformatörünün topraklama sisteminin kontrolüne dayanır. Arıza sırasında güç transformatörünün nötr topraklamasını ayırmaktır. Mesafe rölesinin açılmasındaki hataları önlemek için yeni aşırı akım rölesinin topraklama sistemine eklenmesini önermiştir. Literatürde yapılan çalışmalar incelendiğinde empedansa dayalı yöntemlerde arıza ve arıza yeri tespiti sıfır bileşen akımı, gerilim dalgalanmaları, yüksek empedanslı arızalar ve simetrik olmayan iletim hatlarından dolayı doğru tahmin edilememektedir. Yürüyen dalga tabanlı yöntemler incelendiğinde gerilim ve akım bilgilerini kullanıldığında akım transformatörlerin doyuma gitmesi, yanlış ölçüm sonuçları, sıfır bileşen akımı ve simetrik olmayan iletim hatlarında arıza ve arıza yeri tespiti doğru tahmin edilememektedir. Sezgisel yöntemler incelendiğinde yapılan çalışmalarda gerilim ve akım bilgilerini kullanılarak tahmin çalışmaları yapılmıştır ve akım transformatörlerin doyuma gitmesi, yanlış ölçüm sonuçları, sıfır bileşen akımı ve simetrik olmayan iletim hatlarında gerilim ve akım bilgileri doğru olmadığında arıza ve arıza yeri tespitinde hatalar meydana getirmektedir. Bu makalede yapılan çalışmada arıza anında oluşan empedans diyagramında oluşan görüntüler kayıt altına alınarak görüntü işleme teknikleri kullanılarak arızalar meydana geldiğinde akım transformatörlerin doyuma gitmesi, yanlış ölçüm sonuçları, sıfır bileşen akımı ve simetrik olmayan iletim hatları gibi sorunlarda yüksek doğrulukla arıza ve arıza yeri tahmini yapılmıştır.

Bu çalışmada ikinci bölümde transformatör topraklamasının mesafe koruma rölesi üzerindeki etkisi ve farklı topraklanmış sistemlerde oluşan gerilim ve kısa devre akımlarının incelenmesi yapılmıştır, üçüncü bölümde mesafe koruma rölesinden elde edilen R-X empedans diyagramı görüntüleri ile görüntü işleme teknikleri kullanarak veri seti oluşturulup YSA modeli oluşturularak giriş verisi olarak kullanılmıştır ve arıza yeri tahmini yapılmıştır.

# TRANSFORMATÖR TOPRAKLANMASININ MESAFE KORUMA RÖLESİ ÜZERİNDEKİ ETKİSİ
(TRANSFORMER GROUNDİNG ON DİSTANCE PROTECTİON RELAY)

Transformatör nötr noktası topraklaması bir güç sisteminin tek faz – toprak kısa devre arızası sırasında mesafe koruma rölesinin çalışmasını etkilemektedir. Sistemde oluşan sıfır bileşen akımı iletim hattının hata empedansını etkileyebilir (Meddeb ve diğ., 2019). Topraklama sisteminin arıza üzerinde etkisi üzerinde çeşitli çalışmalar yapılmıştır. Topraklama sistemi ve koruma sistemleri arasındaki ilişki göz önüne alındığında, uygun bir topraklama yapılmalıdır (Meddeb ve diğ., 2019).

Tek faz - toprak kısa devre arıza olması durumunda, güç transformatörleri nötr noktası toprağa bağlı olduğu için sıfır bileşen akımının iletilmesinde çok önemli bir role sahiptir (Lin ve diğ., 2011) ve (Guangfu ve diğ., 2010) . Sıfır bileşen akımı, iletim hattının mesafe rölesi çalışmasını etkilemektedir. Mesafe rölesinin hatalı çalışması güç sisteminin güvenirliğini etkiler.

Transformatörün nötr noktası topraklanmış bir sistemde ve tek faz- toprak hatası sırasında mesafe rölesi tarafından ölçülen empedans, iletim hattının hatalı noktasının empedansıyla orantılı değildir (Meddeb ve diğ., 2019). Bunun sebebi, toprak arızası olayları sırasında sistem topraklamasında sıfır bileşen akımlarının varlığıdır. İletim hattında oluşan sıfır bileşen akımı yüksek frekanslı gerilim dalgalanmalarına ve harmonikleri yol açtığı için bu durum mesafe rölesinin hatalı şekilde çalışmasına bunun sonucunda sistemde oluşan arızaların tespitinde hatalar meydana gelmektedir.

Bu çalışmada güç sistemi için PSCAD ™ / EMTDC ™ benzetim programında modellenen 154kV, 50Hz, 200 km havai iletim hattı tek hat şeması Şekil 1'de gösterilmektedir. İletim hattı üç fazlı iki güç transformatörüne bağlı iki jeneratör üretim kaynağından oluşur. İletim hatları için kabul edilen model, 154kV'luk tek devre enerji iletim hattı 1272 MCM iletkenin kesitleri ve direk tipi 'PB' kullanılmıştır. Transformatörlerin bağlantılarında düşük gerilim tarafı (11kV) üçgen, yüksek gerilim tarafı (154kV) yıldız olarak seçilmiştir. Sistemde 25MW'lık sabit bir yük kullanılmıştır.

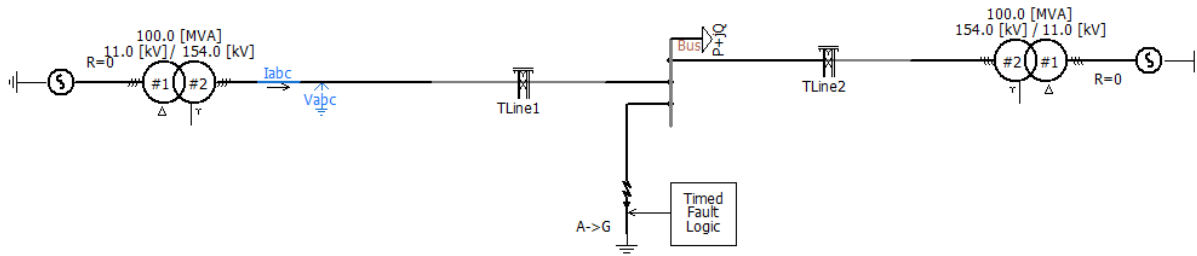

**Şekil 1.** İletim hattı modeli
*Figure 1.* Transmission line model

PSCAD ™ / EMTDC ™  benzetim programında iletim hattı modellenerek tek faz- toprak arızası sırasında transformatör topraklamasının etkisi gösterilmiştir.  Yapılan çalışmada üç farklı şekilde topraklanmamış, doğrudan topraklanmış ve empedansla topraklanmış transformatör sistemleri kullanılarak iletim hattında oluşan kısa devre arızalarının durumu gözlenmiştir. Hata empedansı sabit $Z_f = 1\Omega$ olarak alınmıştır. Kısa devre arızası sabit sürelerde sistem çalıştıktan sonra 0.3 saniye sonra ve 0.05 saniye boyunca devam etmektedir. Çalışmalar en sık karşılaştığımız tek faz-toprak kısa devresi ve diğer fazlarla benzer özellik gösteren (a-fazı) oluşturularak yapılmıştır.

**Topraklanmamış Sistem** (Ungrounded System)

Topraklanmamış sistemlerde güç transformatörleri ile toprak arasında bağlantı yoktur. Bu nedenle, tek fazdan toprağa arıza akımı düşüktür ve sisteme sorun oluşturmazlar. Nötr noktası sıfır bileşen akımının akmasına izin vermez (Meddeb ve ark., 2019). Kısa devre anında Şekil 2'de gösterilen gerilim-akım grafiğinde kısa devre olmayan fazlarda yani sağlam fazlarda faz-toprak aşırı gerilimler oluşmuştur. Kısa devre anında kısa devre yapılan A-fazında yüksek arıza akımı oluşmuştur ve sağlam fazlarda akım artışı olmuştur. Bu tip nötr noktası topraklanmamış sistemlerde arıza akımı çok yüksek olmamasına rağmen yüksek gerilimler oluşması iletim hattı için istenmeyen durumdur.

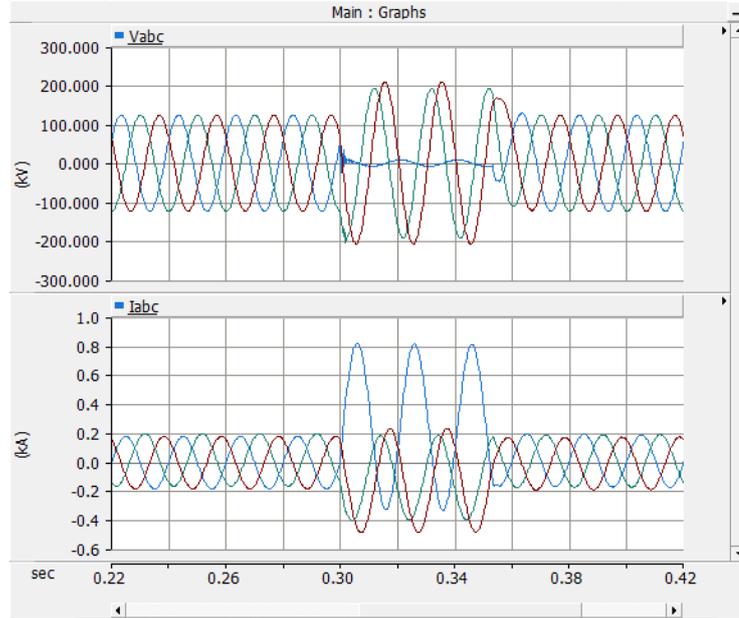

**Şekil 2.** Topraklanmamış sistemde kısa devre anında oluşan gerilim-akım grafiği
*Figure 2. Voltage-current graph appearing during short circuit in ungrounded system*

**Doğrudan Topraklanmış Sistem** (Directly Grounded System)

Doğrudan topraklanmış sistemlerde transformatörün nötr noktası doğrudan topraklanır. Şekil 3'de gösterilen gerilim akım grafiğinde doğrudan topraklanmış sistemin çok yüksek seviyelerde arıza akımı ve kısa devre yapılan hatta gerilim düşmesi gözlemlenmektedir. Sağlam fazlarda gerilim değişmemektedir. Bu nedenle, bu tip sistem topraklaması geçici aşırı gerilimlerin kontrolünü sağlar, ancak yüksek seviyede bir arıza akımı oluştururlar.

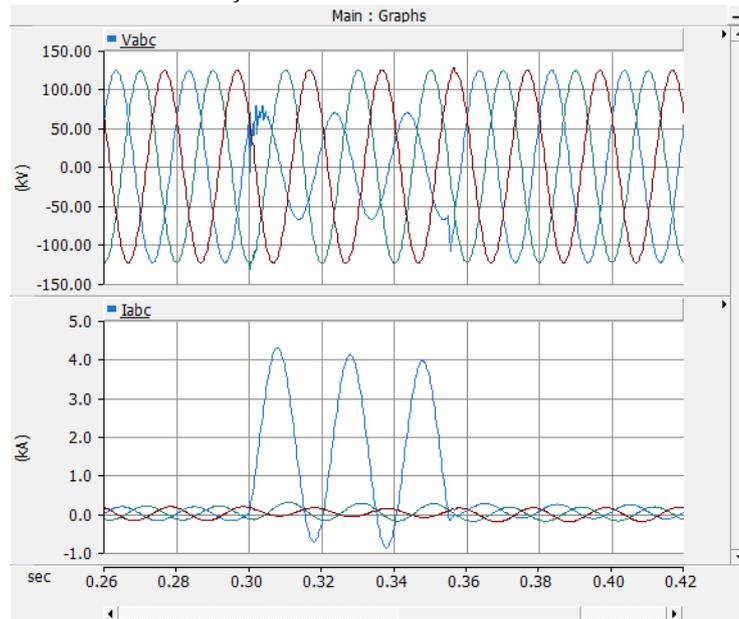

**Şekil 3.** Doğrudan topraklanmış sistemde kısa devre anında oluşan gerilim-akım grafiği
*Figure 3. Voltage-current graph appearing during short circuit in a directly grounded system*

**Empedans ile Topraklanmış Sistem** (Grounded System with Impedance)

Empedans ile topraklanmış sistemlerde transformatörün nötr noktası bir empedans ile topraklanır. Empedans değeri R=5 ohm olarak seçilmiştir. Şekil 4'de gösterilen gerilim-akım grafiğinde empedans ile topraklanmış sistemin arıza akımının, doğrudan topraklanmış sisteme göre empedans nedeniyle daha az olduğunu görülmüştür. Sağlam fazlarda gerilim değişmemektedir ve geçici aşırı gerilimlerin kontrolü sağlar.

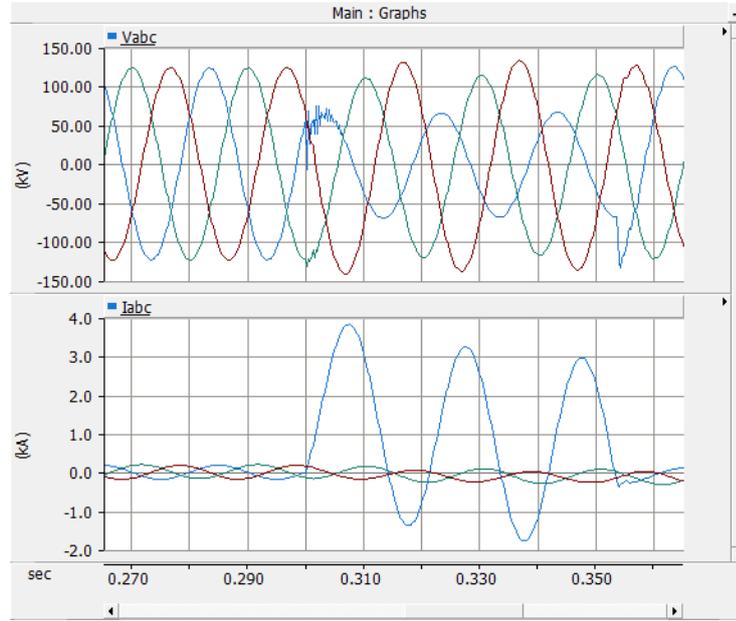

**Şekil 4.** Empedans ile topraklanmış sistemde kısa devre anında oluşan gerilim-akım grafiği
*Figure 4. Voltage-current graph appearing during short circuit in impedance grounded system*

Yapılan çalışmada transformatör nötr noktası doğrudan toprağa bağlandığında, geçici aşırı gerilimlerin oluşmadığı ve yüksek arıza akımı oluştuğu için hata koruma rölesi tarafından hızlıca tespit edilebilir. Ancak, çok yüksek hata akımına sebep olur. Topraklanmamış sistemlerde, arıza akımı düşüktür, fakat iletim hattında önemli yüksek gerilim dalgalanmaları oluşturur. Oluşan yüksek gerilim, iletim hattı donanımlarında ciddi zararlara yol açabilir. Empedans ile topraklanmış sistemlerinin geçici gerilimin ve arıza akımının büyüklüğünü sınırlandırması ve böylece donanımların hasar görmemesini sağlar. Ancak topraklanmış sistem kuruluysa, yatırım pahalı olacaktır (Meddeb ve ark., 2019). Bu nedenle, topraklama tipi, tek fazdan toprağa arıza akımının seviyesini etkiler ve gerilim dalgalanmalarına neden olmaktadır.

**MATERYAL ve METOT** (MATERIAL and METHOD)

Bu çalışmada transformatör topraklamasının mesafe koruma rölesi üzerindeki etkisi incelenerek dengesiz kısa devre arızalarında sıfır bileşen akımı iletim hattında yüksek frekans ve harmonikler oluşturması mesafe koruma rölesinin doğru çalışmasını etkilemektedir. Mesafe koruma rölesinin çalışmasının etkisi azaltmak için yeni bir yöntem üzerinde durulmuştur.

**Görüntü İşleme Tekniği Kullanarak Veri Seti Oluşturulması** (Creating a Data Set Using Image Processing Technique)

İletim hatlarında güç transformatörün topraklama tipine göre farklı kısa devre arızalarına maruz kalmaktadır. Mesafe koruma rölesi kısa devre arızalarında hem arızayı hem de arızanın yerini tespit edebilmektedir. Fakat topraklama tipine bağlı olarak sıfır bileşen akımların ve gerilim dalgalanmaların oluşması mesafe koruma rölesinin arızayı ve arıza yerini doğru tahmin etmesini zorlaştırmaktadır. Bu yüzden bu çalışmada YSA ile arıza tespiti üzerine tahmin çalışmaları yapılmıştır. PSCAD ™ / EMTDC ™ benzetim programında iletim hattı modellenerek mesafe koruma rölesinden elde edilen R-X empedans diyagramı görüntüleri oluşturulmuştur. Şekil 5'de örnek olarak mesafe koruma rölesinden elde edilen R-X empedas diyagramı görüntüleri gösterilmiştir.

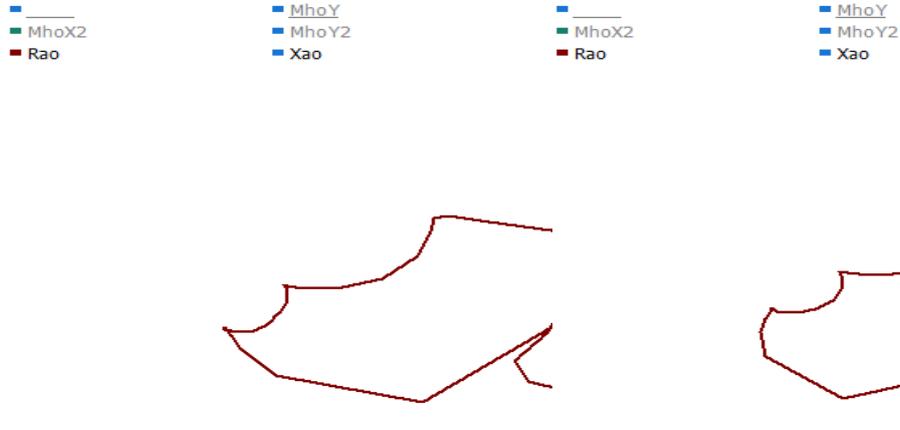

**Şekil 5.** Mesafe koruma rölesi kısa devre anında R-X empedans diyagramı görüntüleri
*Figure 5. R-X impedance diagram images during distance protection relay short circuit*

Bu çalışmada farklı kilometrelerde ve farklı topraklanmış sistemlerde oluşturulan her arızanın R-X empedans diyagramında görüntüleri farklıdır. Elde edilen görüntülerin sayısallaştırılması için Matlab Image Processing Toolbox [MATLAB] programı kullanılmıştır. Her bir görüntü sayısallaştırılarak matris haline getirilmiştir. Sayısal görüntüler renk tonuna göre 0-255 arası değerlerdir, bu değerler normalizasyon yapılarak 0-1 aralığına dönüştürülmüştür. Her bir görüntü, görüntü işleme tekniği kullanılarak sayısallaştırıldığında, 339x292 büyük boyutta bir matris formu elde edilir. Bu boyuttaki bir matris yapay sinir ağlarında eğitilirken işlem süresi uzun ve doğruluk değeri azdır. Bu sebeple istatistiki yöntemler uygulanarak matris boyutu küçültülmüştür. Bu işlem ile YSA uygulanmadan önce kullanılan istatistiki yöntemler ise; aritmetik ortalama ve standart sapmadır. Şekil 6'da görüntülerin elde edilmesi, görüntünün sayısallaştırılması, veri setinin oluşturulması, YSA'ya giriş olarak verilmesi ve YSA'dan çıkış olarak arıza yeri tahminini gösteren akış şeması verilmiştir.

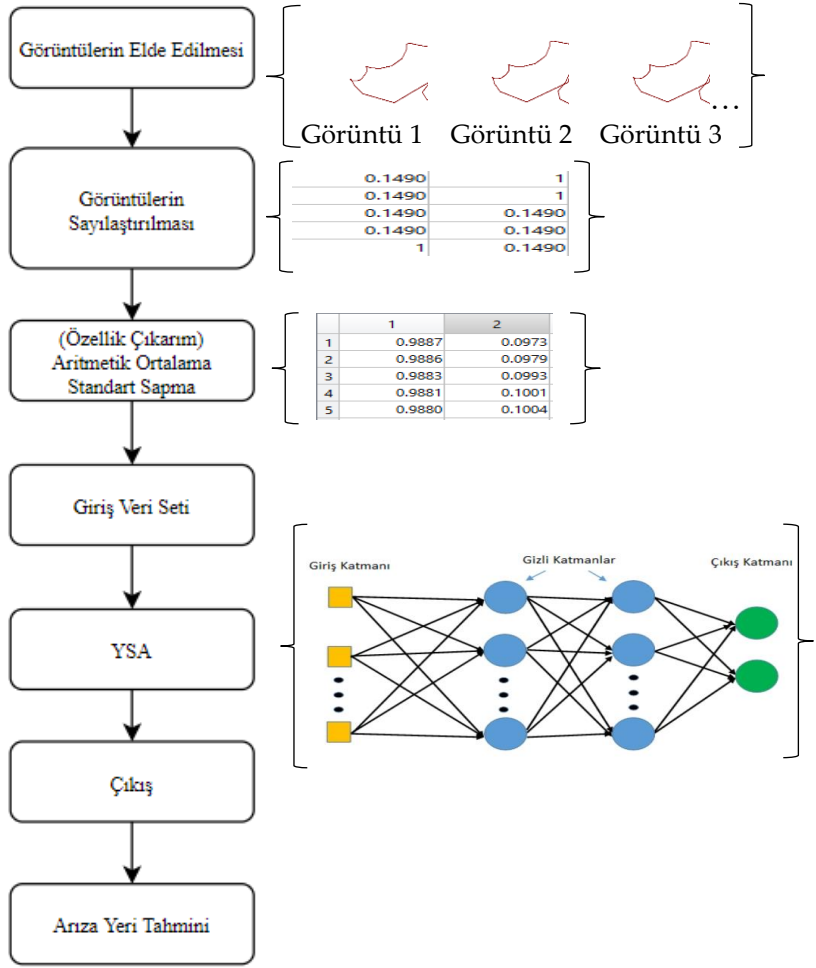

**Şekil 6.** Görüntü işleme teknikleri kullanılarak YSA modeli akış şeması
*Figure 6. ANN model flow diagram using image processing techniques*

**Yapay Sinir Ağı Kullanarak Arıza Yeri Tespiti** (Fault Location Detection Using Artificial Neural Network)

YSA insan sinir sisteminin işleyiş yapısını modelleme fikrine dayalı olarak ortaya çıkan ve çok sayıda sinir hücresinin birbirine bağlanarak oluşturduğu öğrenme şeklini referans alan yapay zeka yöntemlerinden birisidir. Bunun yanında sınıflandırma, tahmin yapma, kestirim gibi problemlerde geniş olarak kullanılan etkili bir öğrenme metodudur.

Bir yapay sinir ağında, giriş katmanı (input layer), gizli katman (hidden layer) ve çıkış katmanı (output layer) olmak üzere temelde üç katmanda oluşur. İlk katman giriş katmanını oluşturur ve dışarıdan girilen verilerin ağırlık değerleri ile çarpılarak gizli katmana iletilmesiyle görevlidir. Bu verilerin istatistikteki karşılığı bağımsız değişkenlerdir. Giriş katmanı probleme etki eden parametrelerden oluşmaktadır. Son katman çıkış katmanıdır ve bilgilerin çıkış olarak aktarılmasını sağlar. Çıkış değişkenlerinin istatistikteki karşılığı bağımlı değişkenlerdir. Modeldeki diğer katman veya katmanlar ise giriş katmanı ile çıkış katmanı arasında yer alır ve gizli katman olarak adlandırılır. Gizli katmanda bulunan nöronların dış ortamla herhangi bir doğrudan bağlantısı bulunmaz. Yalnızca giriş katmanından gelen sinyalleri alırlar ve çıkış katmanına sinyal gönderirler. Gizli katman sayısı ve gizli katmanlarda yer alacak nöronların sayısının seçimi, nöronlara ait aktivasyon fonksiyonun belirlenmesi gibi parametreler oluşturulan ağın performansı açısından önemlidir. Gizli katma sayısı ve nöron sayısı oluşturulacak ağın performansına göre deneme yanılma yöntemi ile belirlenebilir. Temel olarak insan beyninin çalışma şeklini taklit eden yapay sinir ağları veriden öğrenebilme, genelleme yapabilme, çok sayıda değişkenle çalışabilme gibi birçok önemli özelliğe sahiptir (Kırbaş, 2018). Şekil 7'de ileri beslemeli çok katmanlı bir yapay sinir ağı yapısı verilmiştir (Şalvarcı, 2017).

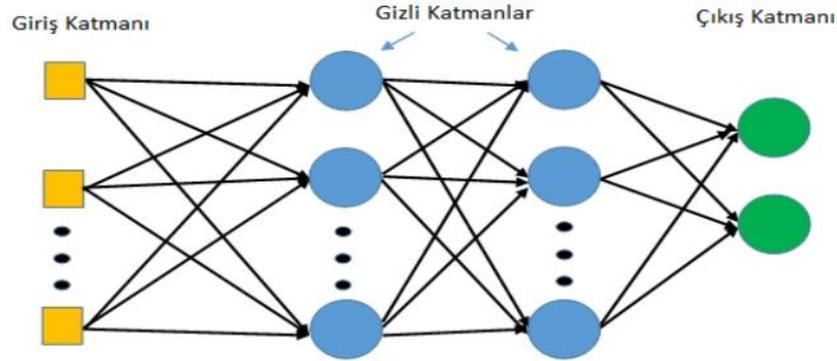

**Şekil 7.** Çok katmanlı ileri beslemeli yapay sinir ağ yapısı
*Figure 7. Multilayer feed forward neural network structure*

Arıza yeri tahmininin YSA yaklaşımı ile belirlenmesi için ağa sunulacak olan veri kümesi görüntü işleme tekniği kullanılarak elde edilen verilerdir. Modelin giriş parametreleri olarak görüntülerden elde edilen sayısal veriler kullanılıp, çıktı olarak her görüntünün arıza yeri 'D_Min_Max' yöntemiyle normalizasyon yapılarak kullanılmıştır. Böylelikle iletim hatlarında arıza yeri tespiti modeli, yapay sinir ağları ile tahmin edilebilmektedir.

YSA çıkışları olarak literatürde yapılan normalizasyon çalışmaları ile ilgili değerlendirmeler incelendiğinde 'D_Min_Max' Normalizasyon yönteminin uygun olduğu tespit edilmiştir.' D_Min_Max' normalizasyon yöntemi için Denklem 1 kullanılır (Yavuz ve Deveci, 2012).

$$X' = 0.8 * \frac{X_i - X_{min}}{X_{max} - X_{min}} + 0.1 \tag{1}$$

Bu eşitlikte; X'= Normalize edilmiş veriyi,
$X_i$= Giriş değeri,
$X_{min}$= Giriş seti içinde yer alan en küçük sayıyı,
$X_{max}$= Giriş seti içinde yer alan en büyük sayıyı
İfade etmektedir.

Tahmin etmeye ilişkin iki önemli adım bulunmaktadır; birincisi veriyi tahmin etmek için hazırlamadır. İkincisi ise farklı tahmin edici modellerin karşılaştırılmasıdır. Modelleri karşılaştırma ölçütleri; doğruluk, hız, sağlamlık, ölçeklenebilirlik, yorumlanabilirliktir. Yapay Sinir Ağları ve makine öğrenmesi yöntemlerinin performans değerlendirmelerinde kullanılan temel performans göstergesi MSE (mean square error) ve yüzde hata değeri kullanılmıştır. Denklem 2 ve Denklem 3'de sırasıyla yüzde hata değeri, MSE hesaplamaları yer almaktadır. MSE ve yüzde hata değeri sıfıra yaklaşmasıyla hata oranının azaldığını göstermektedir (Karasu ve diğ, 2018).

$$\% \, Hata = \frac{Gerçek\ arıza\ yeri - Hesaplanan\ arıza\ yeri}{Hattın\ toplam\ uzunluğu} * 100 \tag{2}$$

$$MSE = \frac{1}{n}\sum_{i=1}^{n}(X_i - Y_i)^2 \tag{3}$$

Bu eşitlikte;
$X_i$= Gerçek arıza yerini,
$Y_i$= Tahmin edilen arıza yerini
İfade etmektedir.

YSA modeli oluşturulurken ağ tipi, eğitim fonksiyonu, gizli katman sayısı, nöron sayısı, transfer foksiyonu YSA performansında çok etkili olduğu bilinmektedir. Çalışmada ağ tipi olarak Feed Forward

Backprop, Cascade Forward Backprop ve Elman Backprop eğitim için kullanılmıştır. Eğitim fonksiyonu olarak Levenberg-Marquardt (LM), Conjugate Gradient with Powell/Beale Restarts (CGB), One Step Secant (OSS), Variable Learning Rate Backpropagation (GDX), Scaled Conjugate Gradient (SCG) kullanılmıştır. Gizli düğüm sayısı için deneme yanılma yöntemi kullanarak en iyi sonucu veren 4 gizli katmanlı YSA modeli oluşturulmuştur. Gizli düğümlerde ki nöron sayısı sırasıyla 20, 18, 10, 5 olarak belirlenmiştir. Öğrenme oranının 0.1' den 5.0' a adımsal olarak artırılması sonucunda 0.9 değerinin en düşük eğitim ve test hatasına sahip olduğu belirlenmiştir. İterasyon sayısının ise 1000 için en iyi sonucu verdiği görülmüştür.

YSA'nın performansını test etmek ve karşılaştırmak için literatürde sıkça kullandığımız makine öğrenmesi yöntemi Destek Vektör Makineleri (DVM) kullanılmıştır. Matlab Regression Learner literatürde yaygın olarak kullanılan makine öğrenmesi yöntemlerini eğiterek karşılaştırma yapma ve en uygun modeli belirleme imkânı sunan bir uygulamadır (MathWorks). Giriş ve çıkış veri setleri uygulamaya verilerek veriler kontrol edilir, özellikler seçilir, doğrulama şemaları belirlenir ve farklı modelleri eğiterek performans sonuçları elde edilir. Her eğitim modelinin sonuçları görselleştirilebilir, yanıt grafiğine, tahmin ve gerçek yanıt grafiğine ve artık grafiğine ulaşılabilir. Eğitim modellerinin tahmini doğruluğunu incelemek ve değerlendirmek için performans ölçütleri kullanılır. Çizelge 1'de Cascade-forward neural network ile DVM eğitim modeli performans ölçütü MSE ile karşılaştırılması gösterilmiştir.

**Çizelge 1.** Cascade-forward neural network ve Destek Vektör Makineleri eğitim modelleri karşılaştırılması

*Table 1. Comparison of Cascade-forward neural network and Support Vector Machines training models*

| Kullanılan Yöntemler | Cascade-forward neural network | Destek Vektör Makineleri |
|---|---|---|
| **Topraklanmamış Sistem (MSE)** | 1,53E-05 | 0,003257 |
| **Doğrudan Topraklanmış Sistem (MSE)** | 4,72E-05 | 0,003460 |
| **Empedans ile Topraklanmış Sistem (MSE)** | 6,61E-05 | 0,001159 |

Çizelge 1'de karşılaştırma sonuçlarına göre kullanılan sistemler değişmesine rağmen Cascade-forward neural network eğitim modelinin MSE performans ölçütü daha iyi sonuç vermiştir.

YSA Cascade-forward neural network eğitim sonucunda Çizelge 2, Çizelge 3 ve Çizelge 4'de gerçek arıza yeri, tahmini arıza yeri, her arıza yerinde yüzde hata miktarı ve MSE sayısal verileri verilmiştir. Yapılan çalışmada minimum hata oranını veren ağ tipi, eğitim fonksiyonu dikkate alınmıştır. Çizelge 2'de transformatörün nötr noktası topraklanmamış olarak elde edilen görüntülerin veri setleri ile oluşturulan YSA modeli, Çizelge 3' de transformatörün nötr noktası doğrudan topraklanmış olarak elde edilen görüntülerin veri setleri ile oluşturulan YSA modeli, Çizelge 4' de transformatörün nötr noktası empedans ile topraklanmış olarak elde edilen görüntülerin veri setleri ile oluşturulan YSA modeli aşağıdaki gibidir. İletim hattında oluşturulan 5-200 km arası 40 kısa devre arızasından 34 tanesi eğitim için 6 tanesi test için seçilmiştir. Şekil 8, Şekil 9 ve Şekil 10'da sırasıyla yapılan çalışmaların gerçek ve tahmin edilen arıza yerleri grafiği gösterilmiştir.

**Çizelge 2.** Topraklanmamış sistemde tahmin edilen arıza yerleri ve hata oranları
*Table 2. Estimated failure locations and error rates in an ungrounded system*

| Kullanılan Metot ve MSE | Ağ tipi: Cascade-forward neural network, Eğitim fonksiyonu: Traincgb, MSE: 1,53E-05 | | | | | |
|---|---|---|---|---|---|---|
| Gerçek Mesafe (km) | 175 | 180 | 185 | 190 | 195 | 200 |
| Tahmin Edilen Mesafe (km) | 174,4020 | 179,3199 | 185,0977 | 191,9809 | 194,1694 | 199,7494 |
| % Yüzde Hata | 0,3066 | 0,3487 | 0,0501 | 1,0158 | 0,4259 | 0,1285 |

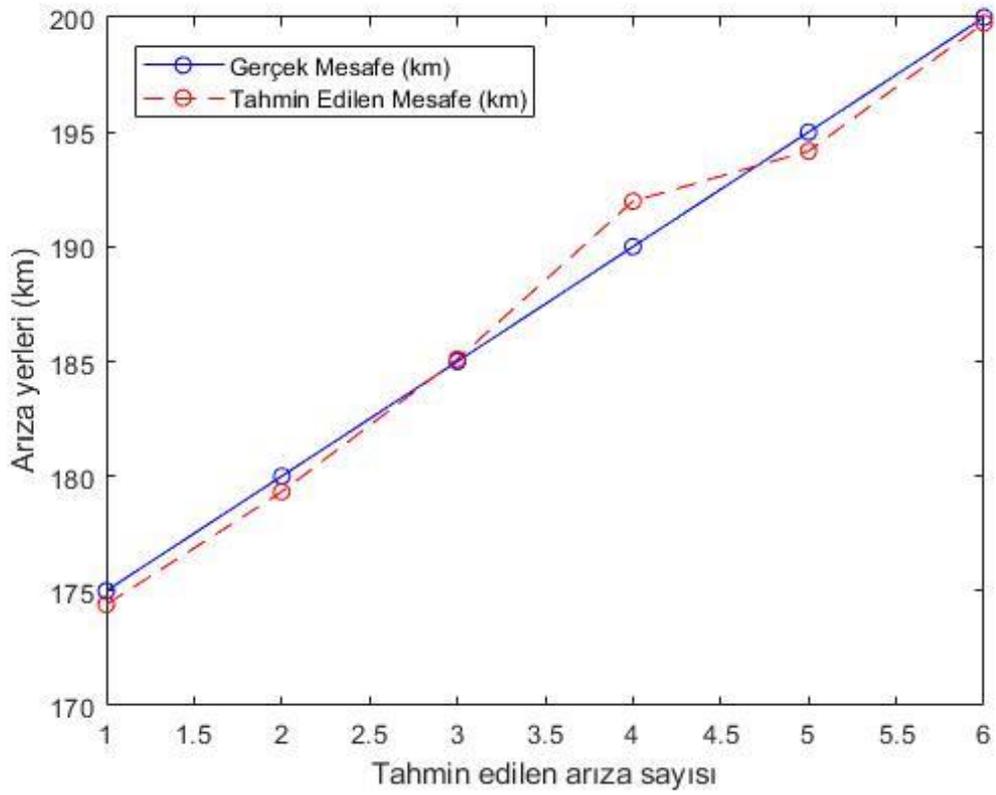

**Şekil 8.** Topraklanmamış sistemde gerçek ve tahmin edilen arıza yerleri grafiği
*Figure 8. Real and predicted fault locations graph in an ungrounded system*

**Çizelge 3.** Doğrudan topraklanmış sistemde tahmin edilen arıza yerleri ve hata oranları
*Table 3. Estimated fault locations and error rates in directly grounded system*

| Kullanılan Metot ve MSE | Ağ tipi: Cascade-forward neural network, Eğitim fonksiyonu: Trainlm MSE: 4,72E-05 | | | | | |
|---|---|---|---|---|---|---|
| Gerçek Mesafe (km) | 175 | 180 | 185 | 190 | 195 | 200 |
| Tahmin Edilen Mesafe (km) | 173,7079 | 178,2277 | 186,8426 | 192,5089 | 196,1746 | 201,0338 |
| % Yüzde Hata | 0,6625 | 0,9088 | 0,9449 | 1,2866 | 0,6023 | 0,5301 |

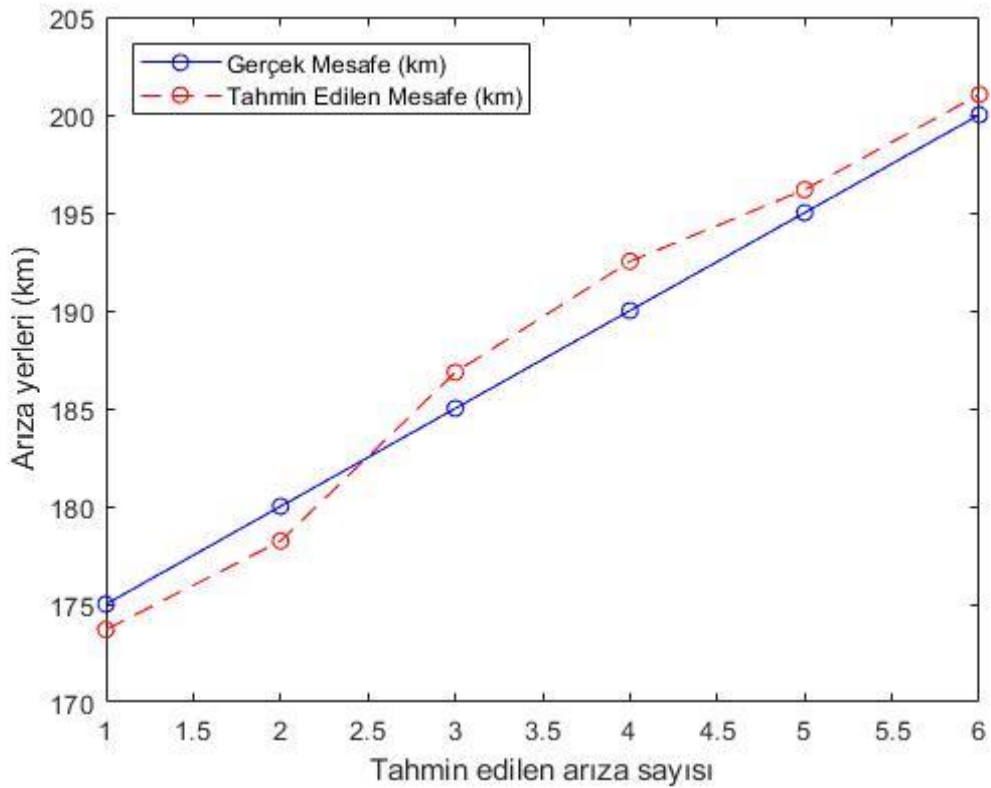

**Şekil 9.** Doğrudan topraklanmış sistemde gerçek ve tahmin edilen arıza yerleri grafiği
*Figure 9. Real and predicted fault locations graph in a directly grounded system*

**Çizelge 4.** Empedans ile topraklanmış sistemde tahmin edilen arıza yerleri ve hata oranları
*Table 4. Estimated failure locations and error rates in an impedance grounded system*

| Kullanılan Metot ve MSE | Ağ tipi: Cascade-forward neural network, Eğitim fonksiyonu: Trainlm MSE: 6,61E-05 | | | | | |
|---|---|---|---|---|---|---|
| Gerçek Mesafe (km) | 175 | 180 | 185 | 190 | 195 | 200 |
| Tahmin Edilen Mesafe (km) | 174,4736 | 178,4958 | 182,7884 | 188,2539 | 193,3469 | 196,7879 |
| % Yüzde Hata | 0,2699 | 0,7713 | 1,1341 | 0,8953 | 0,8476 | 1,6471 |

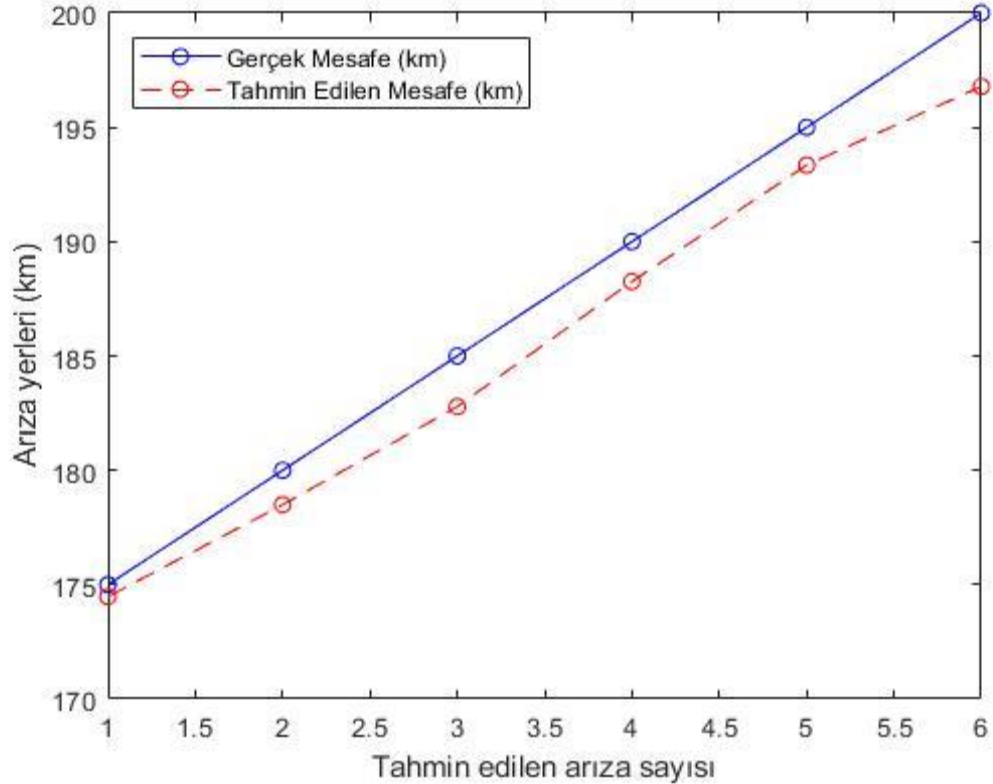

**Şekil 10.** Empedans ile topraklanmış sistemde gerçek ve tahmin edilen arıza yerleri grafiği
*Figure 10. Real and predicted fault locations graph in impednace grounnded system*

**ARAŞTIRMA SONUÇLARI ve TARTIŞMA (RESEARCH RESULTS and DISCUSSION)**

Görüntü işleme teknikleri kullanılarak YSA modeli ile iletim hatlarında farklı yerlerde oluşturulan ve farklı topraklama tipinde yapılan çalışmada arıza yeri tespiti yapılmıştır. Çizelge 2'de elde edilen tahmini arıza yerlerinde en yüksek hata %1,01 ve MSE değeri 1,53E-05 olmuştur. Çizelge 3'de elde edilen tahmini arıza yerlerinde en yüksek hata %1,2866 ve MSE değeri 4,72E-05 olmuştur. Çizelge 4'de elde edilen tahmini arıza yerlerinde en yüksek hata %1,6471 ve MSE değeri 6,61E-05 olmuştur. Bu sonuçlara göre algoritmaların doğruluğu oldukça iyidir ve topraklama sisteminin farklı olmasından etkilenmemektedir.

**SONUÇLAR (RESULTS)**

Bu çalışmada güç transformatörün nötr noktası topraklanmamış, doğrudan topraklanmış ve empedans ile topraklanmış sistemlerin kısa devre arızasında mesafe rölesinin arıza tespit doğruluğunu artıracak çalışmalar yapılmıştır. Yüksek gerilim enerji iletim hatlarında oluşabilecek kısa devre arızaların farklı topraklama tiplerinde arıza yerinin doğru bir şekilde belirlenmesi için YSA tabanlı akıllı sistem tasarlanmıştır. Bu yeni sistemde, YSA girişleri olarak mesafe koruma rölesinden R-X empedans diyagramından alınan görüntüler kullanılmıştır. Yapılan çalışma, PSCAD ™ / EMTDC ™ ile modellenen sistemde test edilmiştir.

YSA kullanımına dayanan arıza ve arıza yeri belirleme algoritmasında kullanılmak üzere belirlenen sinir ağı yapıları denenmiştir ve en iyi olan ağ tipi ve eğitim algoritması kullanılmıştır. Ayrıca YSA'nın performansını test etmek makine öğrenmesi yöntemi DVM ile karşılaştırılmıştır. Çizelge 2, Çizelge 3 ve Çizelge 4 incelendiğinde ve Şekil 8, Şekil 9 ve Şekil 10 karşılaştırma grafikleri görüldüğü üzere yapılan arıza ve arıza yeri belirleme çalışmasında elde edilen hataların düşük olduğu görülmüştür ve algoritmanın iyi bir performansa sahip olduğunu göstermektedir. Tespit edilen arıza mesafe yerleri bize havai hatlarda gözle görülebilecek durumdadır ve arızayı gidermek için yapılan çalışmalarda büyük kolaylıklar sağlamaktadır.

Bu çalışmanın diğer çalışmalardan avantajı ek maliyet, ek cihaz ve karışık matematiksel denklemler olmadan hızlı bir şekilde mesafe koruma rölesinden aldığımız görüntülerin işlenerek YSA tabanlı akıllı bir sistem tasarlanmasıdır.

**KAYNAKLAR** (REFERENCES)